# Real-Time Fatigue Crack Growth Monitoring Using High-Precision Control and Data Acquisition Systems

Arev Hambardzumyan, Rafayel Ghasabyan

TACTUN INC, USA

Emails: arevh@tactun.com, rafayelg@tactun.com

*Abstract* — Fatigue cracks may initiate and propagate long before a structural component reaches the end of its nominal life. Detecting and quantifying crack growth in real time is critical for avoiding catastrophic failures in aerospace structures, nuclear reactors and civil infrastructure. This research reviews four widely used crack-growth monitoring techniques: digital image correlation (DIC), acoustic emission (AE), compliance-based methods and direct current potential drop (DCPD). The research evaluates their working principles, instrumentation, detection thresholds, noise sensitivities and suitability under various environmental conditions. Recent advances in high-precision control and data acquisition systems (DAQs) that enable multi-sensor data fusion are explored. The goal is to guide selection and integration of monitoring methods in modern structural health-monitoring architectures.

## I. Introduction

Structural components operating under cyclic loads accumulate damage in the form of micro-cracks that eventually coalesce into macroscopic fatigue cracks. Without timely intervention, these cracks can propagate rapidly and lead to unexpected failures. Traditional inspections provide only snapshot assessments, whereas real-time monitoring captures crack initiation, crack closure and growth behavior as it occurs. Advances in sensors, high-precision control, wireless networks and data-fusion algorithms have increased the feasibility of continuous crack-growth monitoring. However, selecting an appropriate technique requires understanding each method's capabilities and limitations. The following sections summarize the four principal techniques: DIC, AE, compliance-based methods and DCPD, and discuss how modern DAQ systems allow real-time data integration.

## II. Digital Image Correlation (DIC)

### A. Principle and Instrumentation

DIC is a non-contact optical technique that tracks random speckle patterns on the specimen surface to measure full-field displacements and strains. A high-resolution camera (monochrome or color) images the specimen during cyclic loading; software correlates sequential images to compute displacement fields. The system usually consists of:

- **Speckle pattern**: A random high-contrast pattern applied to the surface using paint or spray to allow sub-pixel correlation.

- **Camera and lens**: High-resolution cameras with suitable lenses record images at a prescribed frame rate. For example, a DIC prototype for aircraft used a 1440×1080-pixel Vision Components IMX296 camera with a 5 mm lens mounted inside a wing section [1].

- **Illumination**: Stable lighting, often via LED rings, minimizes variations in exposure and reduces noise.

- **Processing**: Software divides images into subsets and tracks their movement between frames to compute displacement and strain fields.

### B. Accuracy, Resolution and Detection Thresholds

DIC offers high spatial resolution and displacement accuracy when properly calibrated. A technical note reports that in-plane displacement accuracy is roughly 1/100,000 of the field of view. Typical strain resolution is ≈100 µε, and careful speckle preparation and camera calibration can improve this to ≈20 µε. In a prototype installed in an aircraft wing, images were recorded every 15 s over a 25-hour cyclic test, producing a large data volume, but enabling identification of strain localization and crack propagation [1].

### C. Strengths and Limitations

DIC offers high spatial resolution and displacement accuracy when

- **Full-field measurement**: DIC provides maps of surface strain and displacement, enabling identification of strain concentrations that indicate crack initiation or propagation. It captures crack closure behavior and deformation around the crack tip.

- **Non-contact**: Because the camera is external, there is no influence on specimen stiffness or local stress state.

- **High resolution**: Micro-strain resolution and sub-pixel displacement accuracy allow detection of small cracks and plastic zones.

- **Data volume and computational load**: Recording high-resolution images at a high frame rate produces large datasets. Post-processing or high-performance computing may be necessary, although real-time algorithms are emerging.

- **Line-of-sight requirement**: The technique is limited to accessible surfaces with direct optical access. Out-of-plane motions can cause artefacts in 2D DIC; stereo-DIC or multiple viewing angles mitigate this issue.
- **Environmental sensitivity**: Lighting variations, camera vibration and thermal gradients introduce noise. Field deployment often requires rugged housing and protective lenses.

### D. Applications

DIC is widely used in laboratory fatigue tests on metals, composites and civil structures. It has been applied to detect short cracks in stainless steels and pre-corroded aluminum alloys. In aerospace, a DIC prototype installed inside an aircraft wing monitored crack initiation and propagation; two cameras were used to separate true strain changes from out-of-plane noise. In structural health monitoring of bridges, DIC complements other sensors to validate predicted strain fields.

### III. ACOUSTIC EMISSION (AE)

### A. Principle and Instrumentation

AE monitoring detects transient elastic waves generated by rapid energy releases such as crack initiation, plastic deformation and crack growth. Piezoelectric sensors bonded to the specimen convert these waves into electrical signals. An AE system includes:

- **Sensors and coupling**: Broadband or resonant piezoelectric transducers (e.g., R15α sensor with 50–400 kHz frequency range) are coupled to the structure via coupling gels or adhesives. Multiple sensors enable source localization [2].
- **Signal conditioning**: Preamplifiers (typically 40 dB gain) and band-pass filters (e.g., 100–400 kHz) amplify AE signals while attenuating low-frequency noise [2].
- **Data acquisition**: High-speed ADCs sample at rates of order 1 MHz to capture transient waveforms. Thresholds are set above background noise to trigger data recording.
- **Feature extraction**: AE parameters include amplitude, counts (number of threshold crossings), duration, energy (area under rectified signal), rise time, rise angle, RMS and frequency content.

### B. Detection Thresholds and Sensitivity to Noise

AE systems detect events when the signal amplitude exceeds a preset threshold; typical amplitude thresholds for fatigue crack monitoring are around 50–60 dB, although exact values depend on background noise levels and sensor sensitivity. Because AE sensors pick up any rapid energy release, extraneous noise from friction, impacts or environmental disturbances can trigger false positives. Parameter-based analysis (amplitude, duration, counts) allows rapid processing, whereas multi-parameter analysis incorporating entropy, rise angle and spectral features reduces false positives and better distinguishes crack growth stages.

### C. Strengths and Limitations

- **Early detection**: AE is sensitive to micro-crack initiation and plastic deformation before visible crack growth, allowing preventive maintenance.
- **Internal defect monitoring**: AE waves propagate through bulk material, enabling detection of subsurface damage that optical methods cannot see.
- **Real-time capability**: AE systems provide nearly instant detection because signal processing is triggered by threshold crossings.
- **Noise sensitivity**: Environmental noise can generate false signals; careful sensor placement, shielding and filtering are required. Multiple sensors and source localization can help discriminate real events.
- **Calibration and interpretation**: Absolute AE parameter values depend on sensor type, attenuation and coupling; therefore, empirical or machine-learning models are often used to correlate AE features with crack size

### D. Applications

AE has been used extensively to monitor fatigue crack growth in steels, aluminum alloys and composites. The technique has proven effective in detecting early crack initiation in corrosive environments [3] and in monitoring crack growth rates in Hadfield steel [4]. In industrial applications, AE is often combined with DIC or DCPD to provide complementary information on crack initiation and propagation.

### IV. COMPLIANCE-BASED METHODS

### A. Principle and Instrumentation

Compliance-based methods infer crack length from the specimen's compliance (inverse stiffness) under cyclic loading. When a crack grows, the specimen becomes more compliant, causing greater displacement under the same load. Two common approaches are:

- **Back-face strain (BFS) or front-face strain gauges**: Strain gauges measure the opening displacement near the crack mouth; compliance is calculated from the ratio of displacement to applied load.
- **Load–displacement measurements**: Servo-hydraulic test machines record load and crosshead displacement. A finite element model or empirical calibration curve relates compliance changes to crack length

### B. Accuracy, Resolution and Detection Thresholds

The accuracy of compliance methods depends on the quality of the calibration and the measurement resolution of load and displacement. Finite element analysis or analytical solutions (e.g., ASTM E647 equations) are used to derive compliance–crack length relations. In practice, calibration errors, crack shape assumptions and machine compliance influence accuracy. A normalized area-compliance method combining compliance measurement with optical crack length measurement achieved a crack depth resolution of 0.03 mm and an accuracy of 0.08 mm [5]. Typical errors in potential drop crack-depth measurements (an alternative to compliance) are 10–20 %. The compliance method is less sensitive to micro-crack initiation compared with

AE but reliably tracks crack growth once the crack length is within the calibrated range.

C. *Strengths and Limitations*

- **Simplicity and cost-effectiveness**: Only load and displacement sensors are required, and many fatigue testing machines already provide these signals. Automated systems can use closed-loop control to keep the load constant while monitoring compliance.

- **Reliability in standard geometries**: For compact-tension or C-ring specimens, closed-form solutions relate compliance to crack length. The method is widely used in fracture toughness and fatigue crack growth testing (ASTM E647). An automated test system with compliance monitoring provided stable load control and environmental stability, detecting crack extension and crack closure in corrosive environments.

- **Calibration complexity**: Compliance–crack length relations depend on specimen geometry, crack shape and loading conditions. Experiments or finite element simulations are needed for each new geometry. Calibration is complicated when multiple interacting cracks are present, and misalignment or plasticity can introduce errors. A review noted that complicated calibration is one of the reasons compliance methods have limited application in fatigue tests.

- **Resolution limits**: The method is less sensitive to very small crack extensions compared with DIC or DCPD. Strain-gauge noise and machine compliance limit resolution; typical detection thresholds are on the order of a few tenths of a millimeter.

D. *Applications*

Compliance methods are standard for characterizing fatigue crack growth rates in compact-tension specimens, often in conjunction with fracture toughness measurements. They have been incorporated into automated environmental crack-growth systems with finite-element-calibrated compliance curves. Compliance methods are also used in the normalized area-compliance technique for cylindrical rods, where combining compliance with optical measurement provides high accuracy.

## V. Direct Current Potential Drop (DCPD)

A. *Principle and Instrumentation*

DCPD monitors crack growth by measuring the voltage drop across a specimen carrying a constant direct current. As a crack propagates, the effective cross-sectional area reduces, increasing electrical resistance and thus the potential drop. A typical DCPD system comprises:

- **Current source and probes**: A stable current is injected through the specimen, and potential probes (often spot-welded) measure the voltage difference across a defined gauge length.

- **High-precision voltmeter**: Because metals have low resistance, high currents and sensitive differential voltmeters or instrumentation amplifiers are necessary to resolve small voltage changes.

- **Calibration**: Analytical expressions or finite element models relate potential drop to crack length. The gauge length (distance between probes) influences sensitivity: reducing probe spacing increases sensitivity but also increases susceptibility to electronic noise.

B. *Accuracy, Resolution and Detection Thresholds*

The resolution of DCPD depends on crack length and probe spacing. Smaller spacing improves resolution because the potential drop change per unit crack growth is larger, but it reduces the absolute voltage and makes the measurement more susceptible to noise. Calibration is required because analytic solutions often underestimate crack length for curved crack fronts; calibration techniques include finite element simulations, sawn cuts and marker loads. Temperature changes alter resistivity, and electrical bridges formed by contact between crack faces can distort signals; controlling the specimen temperature and using the maximum potential drop per cycle reduce these errors. DCPD is less sensitive to very short cracks compared with AC potential drop methods but has been shown to detect crack initiation in fatigue experiments.

C. *Strengths and Limitations*

- **Robustness in harsh environments**: DCPD systems are stable in high-temperature, high-pressure and electromagnetically noisy environments, making them suitable for monitoring crack growth in nuclear reactor components.

- **Simple instrumentation and automation**: Once probes are attached, current and voltage can be recorded simultaneously with load and displacement data, enabling automated crack growth monitoring.

- **Insensitive to optical constraints**: DCPD can detect internal cracks and does not require line-of-sight.

- **Limited sensitivity to short cracks**: DCPD is less sensitive to very short cracks or multiple small cracks. As cracks initiate at multiple sites, DCPD reflects an average crack length and cannot identify which crack will dominate.

- **Calibration and contact issues**: Measurement accuracy depends on reliable probe attachment and stable electrical contacts. Magnetostriction in ferromagnetic steels (the Villari effect) can alter the potential drop; selecting non-magnetic materials or using AC potential drop methods can mitigate this.

D. *Applications*

DCPD is widely used in fatigue crack growth tests on compact-tension specimens and surface-cracked rods. It is commonly paired with compliance or AE measurements to improve accuracy. In pressurized water reactor (PWR) environments, DCPD has been applied to stainless-steel bars at 300 °C and 150 bar, successfully detecting crack initiation and estimating growth rates [6]. Because of its resistance to electromagnetic interference and temperature stability, DCPD is often the method of choice in nuclear and high-temperature applications.

## VI. High-Precision Control and Data Acquisition Systems

### A. Sensing Skin and Large-Area Electronics

Conventional sensors monitor crack growth at discrete locations, leaving large regions uninstrumented. Large-area electronics (LAE) enable wide-area coverage with conformable sensors. An example is the *sensing skin* developed for steel-bridge monitoring using soft elastomeric capacitors (SECs). Each SEC acts as a strain sensor; its capacitance changes with surface strain. To improve sensitivity, researchers corrugated the dielectric to enhance contact with the conductive layers and developed a wireless data-acquisition node based on a De Sauty bridge and a 24-bit ADC integrated into a low-power Xnode smart sensor platform [7]. The DAQ includes a calibration circuit, solar-powered battery, 2.4 GHz radio and optional 4G connectivity. Compared with commercial capacitive DAQs, the custom board achieved 34 % lower noise when monitoring strain under traffic loading. Signal processing uses a continuous wavelet transform to compute a crack growth index (CGI); a threshold corresponding to 30 µε triggers event recording, enabling early detection of crack propagation in the field.

### B. Multi-Sensor Data Fusion

Integrating multiple sensors reduces uncertainty and exploits complementary information. A patented multi-sensor fusion method combines Lamb-wave data from piezoelectric sensors and strain data from conventional strain gauges. The method extracts signal envelopes, removes elastic deformation components and uses a random-forest classifier to distinguish crack states. Dempster–Shafer (D–S) evidence theory fuses the classifier outputs to improve reliability [8]. The workflow divides data into training and testing sets, builds decision trees using the C4.5 algorithm and uses majority voting to classify each feature sample before fusing them with D–S evidence. Such approaches enable real-time crack-growth monitoring even when individual sensor signals are noisy or ambiguous.

In AE monitoring, multi-parameter analysis similarly enhances crack detection. A 2022 study extracted amplitude, energy, entropy, rise time, duration, rise angle, RMS and frequency-domain features from AE signals and correlated them with stress intensity factor range and crack length [2]. The study used DCPD to measure crack length and employed AE parameters to characterize different fatigue stages, demonstrating that combining multiple features reduces false positives and improves early detection. High-precision DAQ systems with 1 MHz sampling rates and 100–400 kHz band-pass filters were essential for capturing these parameters.

### C. Real-Time Control and Closed-Loop Testing

Modern servo-hydraulic or electromechanical test machines incorporate closed-loop controllers capable of maintaining constant load, displacement or stress-intensity factor while simultaneously monitoring crack length via compliance or DCPD. An automated environmental cracking system used finite-element-calibrated compliance curves to adjust actuator displacement and maintain target stress intensity factors, enabling stable crack growth testing under corrosive environments. The system's high-precision control loops provided accurate crack length measurements and allowed detection of crack closure behavior.

In DCPD systems, constant-current sources with feedback control stabilize the injected current, and high-resolution ADCs record voltage drop simultaneously with load and displacement. Data acquisition systems integrate the signals, and software algorithms calibrate potential drop to crack length in real time. Temperature compensation circuits and signal-averaging algorithms mitigate drift.

### D. Challenges and Opportunities

The main challenge in high-precision DAQ integration is synchronizing heterogeneous data streams (images, acoustic waveforms, strains and voltages) and achieving adequate sampling rates for each sensor. Data fusion algorithms must account for different noise characteristics and delays. Emerging edge computing platforms and wireless sensor networks enable on-board processing and reduce transmission load, making real-time multi-sensor monitoring viable for field applications.

### E. Guidelines for Selecting a Monitoring Technique

Selecting an appropriate crack-growth monitoring technique depends on the structural material, environment, crack morphology and desired information. The following guidelines provide a starting point:

1. **Detection of crack initiation and micro-cracks**: AE excels at detecting early micro-crack events and plastic deformation. It should be used when early warning is critical and environmental noise can be controlled or filtered.

2. **Full-field surface strain mapping**: DIC is ideal when detailed strain maps or crack opening measurements are needed. It requires optical access and stable lighting but provides high spatial resolution and can quantify crack-tip plastic zones.

3. **Standardized crack growth measurement**: Compliance-based methods are suited for laboratory testing with standard geometries (compact-tension, C-ring, etc.). They provide reliable crack length measurements after calibration and can be integrated into automated control loops.

4. **Harsh environments and internal cracks**: DCPD is preferred in high-temperature or electromagnetically noisy environments and when optical access is impossible. It provides robust crack length tracking but may miss very short cracks.

5. **Multiple interacting cracks**: Combine DCPD or compliance with AE or DIC to discriminate between crack initiation sites and dominant crack growth. Multi-sensor data fusion using machine-learning algorithms enhances reliability.

6. **Real-time field monitoring**: For large structures such as bridges, distributed sensing skins with wireless DAQ provide continuous strain mapping over wide areas. Data thresholds (e.g., 30 µε) trigger event recording and reduce data volume. These systems should be combined with periodic visual inspections and other NDT methods.

## Conclusion

Real-time monitoring of fatigue crack growth is essential for ensuring the safety and reliability of critical structures. Each technique reviewed: DIC, AE, compliance-based methods and DCPD, offers unique strengths. DIC provides high-resolution surface strain maps but requires optical access. AE detects micro-crack initiation and internal damage but suffers from noise. Compliance-based methods offer simple instrumentation and are widely used in standard tests yet require careful calibration. DCPD excels in harsh environments and provides robust crack-length measurement but is less sensitive to very short cracks. Modern high-precision control and data acquisition systems facilitate multi-sensor integration, enabling complementary techniques to be combined. Examples from aerospace, nuclear power and bridge monitoring demonstrate that integrating sensors, advanced DAQ hardware and data-fusion algorithms can provide early warnings, reduce downtime and extend the life of critical components. Continued research into sensor technologies, edge computing and machine learning will further improve the fidelity and practicality of real-time fatigue crack monitoring.

## References

[1] L. Luan, L. Crosbie, S. Michel, E. Hack, "A digital image correlation (DIC) prototype system for crack propagation monitoring in aircraft assemblies", PMC PubMed Central, 2022.

[2] M. Chai, C. Lai, W. Xu, Q. Duan, Z. Zhang, Y. Song, "Characterization of fatigue crack growth based on acoustic emission multi-parameter analysis", PMC PubMed Central, 2022.

[3] N. Angelopoulos, V. Kappatos, "An experimental assessment using acoustic emission sensors to effectively detect surface deterioration on steel plates", PMC PubMed Central, 2024.

[4] S. Shi, Guiyi Wu, H. Chen, S. Zhang, "Acoustic emission monitoring of fatigue crack growth in Hadfield steel", MDPI, 2023.

[5] C.Q. Cai, C.S. Shin, "A normalized area-compliance method for monitoring surface crack development in a cylindrical rod", International Journal of Fatigue 27, 2005

[6] S. Arrieta, F. J. Perosanz, J. M. Barcala, M. L. Ruiz, Sergio Cicero, "Using direct current potential drop technique to estimate fatigue crack growth rates in solid bar specimens under environmental assisted fatigue in simulated pressurized water reactor conditions", MDPI, 2022.

[7] H. Liu, S. Laflamme, J. Li, A. Downey, C. Bennett, W. Collins, P. Ziehl, H. Jo, M. Todsen, "Sensing Skin Technology for Fatigue Crack Monitoring of Steel Bridges", International Journal of Bridge Engineering, Management and Research, 2024

[8] Xian University of Technology, "A Crack Growth Monitoring Method Based on Multi-sensor Data Fusion", US Patent CN111024527B, 2019